\documentclass{article}
\usepackage[utf8]{inputenc}
\usepackage{authblk}
\usepackage{setspace}
\usepackage[margin=1.25in]{geometry}
\usepackage{graphicx}
\graphicspath{ {./figures/} }
\usepackage{subcaption}
\usepackage{amsmath}
\usepackage{lineno}
\usepackage{hyperref}
\usepackage[T1]{fontenc}
\usepackage{CJKutf8}

\usepackage[style=nejm, 
citestyle=numeric-comp,
sorting=none]{biblatex}
\title{Macroscopic Quantum Resonators Path Finder (MAQRO-PF) White Paper}

\author[1]{Jack Homans}
\author[1]{Laura da Palma Barbara}
\author[1]{Jakub Wardak}
\author[1]{Elliot Simcox}
\author[1]{Tim M. Fuchs}
\author[1*]{Hendrik Ulbricht}

\affil[1]{School of Physics and Astronomy, University of Southampton, SO17 1BJ, Southampton, UK}
\affil[*]{Address correspondence to: H.Ulbricht@soton.ac.uk}

\date{}

\begin{document}

\maketitle

\begin{abstract}
\noindent Optically levitated particles are used in a wide range of experiments to explore both fundamental physics and to act as sensors to a variety of external forces. One field of particular interest that these particles can be used to investigate is quantum mechanics. Previous research has yet to set an absolute upper bound on the size of objects that can be prepared in a quantum superposition. Exploring this limit involves allowing ever-larger objects to freely and coherently evolve to assess if their behaviour matches quantum or classical theoretical predictions. However, the long free evolution times required for these behaviours to be visible result in the experiments being gravitationally limited. Space based platforms therefore become the next key step in these investigations. In this white paper, we shall discuss our proposal for an optical levitation experiment in space that will explore the fundamental upper size limits of quantum mechanics. We shall cover the scientific motivation behind these investigations, then summarize the current status of our designs for the satellite. We will then review the aspects of the payload that require further development, then summarize the current estimates of the payload's requirements.\\\\
This white paper has been submitted into the UK Space Agency's Frontiers 2035 road mapping programme for the UK's future space science. Comments are welcome.
\end{abstract}


\section{Scientific Motivation and Objectives}
We propose a pathfinder (PF) mission on a dedicated satellite in low Earth orbit (LEO) to perform quantum science experiments that will test the limits of quantum mechanics and the interplay between quantum mechanics and gravity. These scientific objectives are defined by the Macroscopic Quantum Resonators (MAQRO) mission proposal which is well-known to ESA \cite{MAQRO2012,MAQRO2015update,MAQRO2023}. The MAQRO mission objectives have been evaluated in the QPPF CDF in 2018 \cite{QPPF_CDF} and are included in space science roadmaps by ESA's Voyage 2050 and NASA's BPS 2023 Decadal Survey. The proposal has also recently undergone a CDF study at the University of Portsmouth in September 2025. Our fundamental science experiment is based on levitated optomechanics, a mature technology that optically traps and controls the motion of silica nanoparticles with masses much larger than an atom. MAQRO-PF is based on the space-qualified Op-to-Space payload which aimed to demonstrate levitated optomechanics technology on a rideshare mission hosted by The Exploration Company in June 2025 \cite{homans2025experimental}. MAQRO-PF will extend the quantum coherence (drag-free) evolution time for macroscopic quantum experiments to at least 10 seconds--an order of magnitude greater than is possible on Earth. The key science requirement is to reach a drag-free environment of $10^{-9}$ g for the payload over such periods. The other requirement current best estimates are stated in Section \ref{Sec_4}.\\
\begin{figure}[b]
    \centering
    \includegraphics[width=\linewidth]{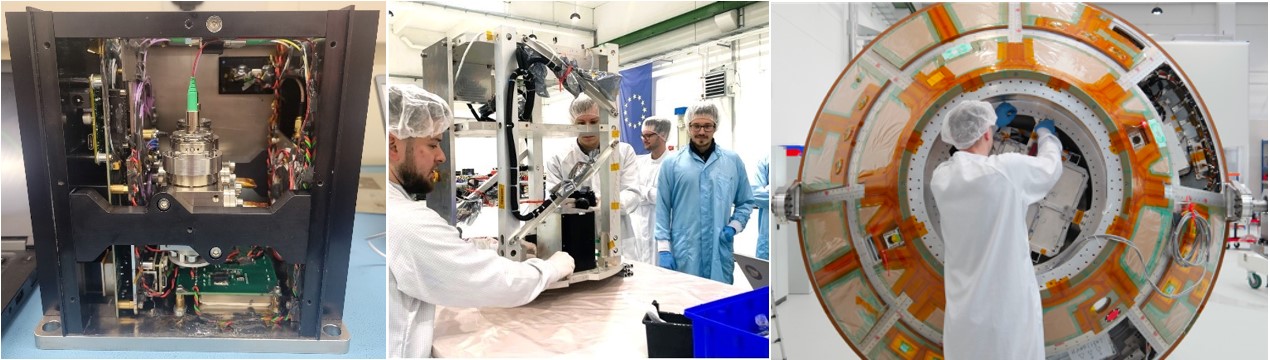}
    \caption{\textbf{[Right]} The Op-to-Space payload being integrated into \textbf{[Left]} The Exploration Company's Nyx `Mission Possible' capsule.}
    \label{OtS_integration}
\end{figure}\\
The \textbf{science goals} of the MAQRO-PF mission are well aligned with the UKSA's in-scope activities of fundamental physics and technology development:
\begin{itemize}
    \item Demonstrate in-orbit technology for large mass quantum superposition in space.
    \item Conduct the first tests of fundamental physics in space: limits of quantum mechanics, the interplay of quantum mechanics and gravity, and Dark Matter \cite{bassi2022way,belenchia2021test,belenchia2022quantum,snellen2022detecting}.
    \item Extract key parameters on in-orbit decoherence and noise effects to inform a future MAQRO mission at Earth-Sun L2.
\end{itemize}
\subsection*{Alignment with UK space strategy, UKRI funding and UKSA agenda}
Our mission's aims are aligned with the UK space strategy's \textbf{Space Science theme} as we aim to explore the fundamental physics behind and improve our understanding of quantum mechanics \cite{millen2020quantum,carlesso2022present,bose2025massive}. The experiments proposed also hold the potential to explore other fundamental fields such as detecting Dark Matter and Dark Energy candidates \cite{moore2021searching, kilian2024dark, amaral2025first, kalia2024ultralight, adelberger2022snowmass} and gravitational waves \cite{arvanitaki2013detecting, pontin2018levitated}. Additionally, they could be used as extremely precise force sensing devices \cite{attonewton_force_sensing,electric_force_sensing,2020_optic_lev_accelerometer, fuchs2024measuring}. The MAQRO proposals represent the interest of a large international research community \cite{gonzalez2021levitodynamics}, and undertaking the mission in partnership with the UKSA would maintain the UK arm of this community at the forefront of the research. This would contribute to the ambition of the Science and Technology Framework--making the UK a science and technology superpower by 2030 \cite{STF}. Additionally, it would provide opportunities for collaboration with both academic and commercial partners inside the UK and on an international scale, in-line with the \textbf{`International Relations and Partnerships' thread} of the UK space strategy \cite{UKSA_corporate_plan}. Furthermore, a wide range of technical development must be completed before the mission reaches maturity, providing opportunities for spin-outs to be created to commercialize the research, both for use inside the research community and in the wider field of satellite technology. The Op-to-Space payload already collaborated with UK startup companies such as Aquark and Twin Paradox Labs in its development and was linked to various members of the Space South Central cluster. Future payload design and construction would open opportunities for further investment in these companies and others, stimulating their growth and satisfying the \textbf{`Innovation, Investment and Commercialization' thread} of the UK space strategy \cite{UK_space_strategy}. Further to this, developing the payload would create openings to attract and train both local and global talent, alongside inspiring new generations into the rapidly growing UK quantum marketplace. This was previously demonstrated with the Op-to-Space payload where primary school students entered art pieces into a competition to have them embossed onto the payload. This growth of the UK talent pool would fulfil the \textbf{`Education and Future Workforce' thread} of the UK space strategy and would link to the goal of building a skilled space workforce, as set out in the Space Industrial Plan \cite{Space_industrial_plan}.
\subsection*{Science Background}
Quantum physics is a mature field with wide-ranging uses and impacts in our daily lives. Interest in quantum technologies in space has grown rapidly but has predominantly focussed on commercial and security applications. One such system is the 2016 Chinese Micius satellite which demonstrated quantum-encrypted communication between Beijing and Vienna \cite{micius}. Another example is the first space-based Bose-Einstein condensate (BEC)–-a quantum system with applications in sensing and metrology–produced by a German team in 2017 \cite{maius}, leading to a collaboration with NASA in their Cold Atom Laboratory aboard the ISS in 2020. However, some areas of quantum physics remain relatively unexplored. One of the remaining questions is where the boundary between quantum and classical descriptions of reality start and end. The key feature of quantum physics lies in the wave-like behaviour of objects. To study the quantum limit, ever heavier objects are shot through narrow slits, where they create distinctive interference patterns in the same way as light and water waves do. The current largest quantum objects are sodium clusters that are thousands of times smaller than bacterium, containing only 5,000–-10,000 atoms, measuring 8 nanometres across and weighing 170,000 atomic mass units\footnote{(atomic mass units = amu; 1 amu is one-twelfth the mass of a carbon-12 atom)} \cite{largest_superposition}. However, despite extensive interest from technological development, further exploration has been stunted by the limits of what can be done in laboratories.\\\\
To produce accurate data, quantum-matter interferometers must shield the quantum system from outside noise sources such as external gases, light and vibration. This often makes them complex and bulky. Larger particles also take longer to exhibit quantum effects as their waveforms spread out more slowly. This means that interactions between larger particles and the environment get more common, washing out the quantum behaviour. These longer experimental periods also result in gravity becoming a limitation as it causes the particles to fall out of tabletop experiments after only a few milliseconds. This constrains our experiments as particles larger than a few tens of nanometres ($>10^{9}$ amu) require upwards of 10 seconds of free evolution for their quantum effects to be distinguishable from classical models. Attempts to keep the particles in the experiment with magnetic or electric fields or lasers only serve to decohere the particle out of its quantum state, preventing useful data from being collected.\\
\begin{figure}[t]
    \centering
    \includegraphics[width=0.5\linewidth]{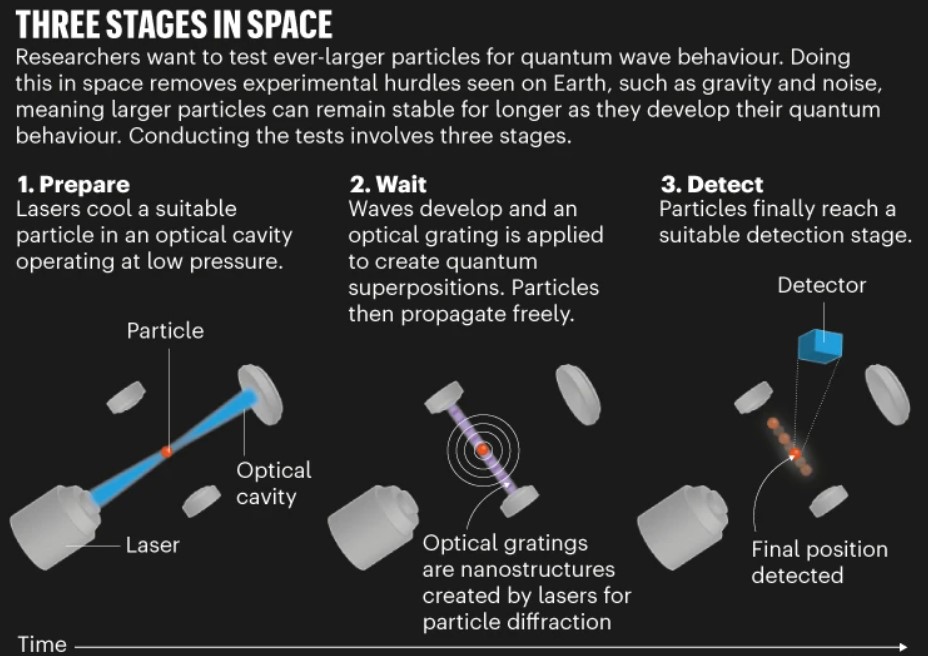}
    \caption{Process of conducting matterwave interferometry in space \cite{belenchia2021test}.}
    \label{matterwave_fig}
\end{figure}\\
The answer is to work in space, where the microgravity conditions would allow particles to float for minutes without leaving the trap. They can fall towards Earth at the same rate as the satellite containing the experiment, facilitating $>100$ second experiments that are equivalent to controlling a particle dropping from 50 kilometres. However, major challenges must still be overcome. The whole setup must be designed for the specific particles being used, accounting for their sizes and electric charges among other key parameters. The payload must be protected from the harsh space environment; cosmic rays, solar winds and ionizing radiation. Size and weight limitations must be accounted for, and the satellite's motional noise–-such as vibrations from propulsion systems–-must be minimized.\\\\
The optical traps used to confine and characterize particles before they are released into their coherent quantum state are formed by using a parabolic mirror to focus laser light down to an approximately 1 $\mu$m $\times$ 1 $\mu$m $\times$ 1 $\mu$m point. These traps are used as they are easy to align, and the parabolic mirrors maximize the amount of light that we collect from the trapped particles, facilitating more precise measurement of the particle's position. However, the small trapping region poses a significant challenge to our experiments, as, during the 10–-100 s period in which the particle is released, the particle and trap cannot move more than 1 $\mu$m relative to each other. Selective release algorithms are under development that will allow us to release the particles when their motion relative to the trap is negligible. However, other factors such as atmospheric drag and solar radiation pressure will cause the satellite, and therefore the trapping region, to accelerate away from the particle. Gravitational variation effects from the Earth (such as J$_2$ perturbations) were also considered as sources of acceleration but were dismissed as they will influence the satellite and the particle identically. ESA's DRAMA (Debris Risk Assessment and Mitigation Analysis) tool was used to assess the effects of orbital perturbations in a worst-case scenario, and found that, for a 100 kg satellite in a 600 km orbit, the atmospheric drag during maximum solar activity would cause the satellite's orbit to decay at a rate 4 orders of magnitude greater than would be acceptable for our experiment. A floating optical bench design has therefore become the main focus of our designs (discussed in Section \ref{Design_concepts_sec}).\\\\
Alongside drag effects, the particle can also be lost from the trap if the vibrational or rotational environment of the host satellite is too large. For linear vibrations, a 1 second release time would require a vibrational stability of \textbf{at least 10\textsuperscript{-7} g}, while a 100 second release time would require \textbf{at least 10\textsuperscript{-9} g} to keep the particle within the 1 $\mu$m$^3$ trap. For platforms such as the ISS, the standard linear vibrational stability sits, at best, in the $\mu$g range, which would not be suitable. Rotations of the payload can also result in the trapping region moving away from the particle if the trap is not positioned exactly on the satellite's centre of rotation (COR). Assuming that the trap is within 1 cm of the satellite's COR, a \textbf{rotational stability of 0.1 mRad / s} would be required. However, this will decrease linearly as the trap's distance from the COR increases. Unless the satellite is designed specifically to carry the optical trapping payload, positioning the trap so close to the COR will be exceedingly difficult. As mentioned above, the floating optical bench design is therefore a required part of the payload such that the payload is decoupled from both the satellite's linear vibrations and rotations while the particle is released.\\\\
As well as needing long periods to release the particle into a quantum state, we must also repeat the experiment many times to build up the data that we need. In any single experiment, the data we collect allows us to measure the particle's position once it is recaptured. By collating the recapture positions of many particles, we can build up a heat-map of where the particles end up. By comparing the heat map with theoretical predictions from both quantum and classical physics, we can determine whether the particle has existed in a quantum state. However, to attain sufficient statistics to explicitly determine the particle's behaviours, we must release and recapture the particle at least 1000 times to create a clear heat map. Furthermore, if the particle is lost from the trap, we must restart with a new particle which must be characterized and then have its own statistical data set collected. Additionally, to maximize the information that can be obtained from the payload, we would aim to include sources of particles of a variety of sizes such that we can explore how large of an object can be maintained in a quantum state. Therefore, we would expect to run experiments on a large number of particles, each requiring at least 1000 experimental procedures to collect sufficient data, meaning that we would hope to run possibly hundreds of thousands of experimental procedures. With an expected \textbf{300 experiments per day}, the goal is thus to put our payload into space for long periods to take all the required data.

\section{Design Concepts}
\label{Design_concepts_sec}
\subsection*{Form Factor}
Two main design concepts exist for our mission. The first has a smaller form factor \textbf{(10--20 U)} and could be designed to fit as a ride-share or a payload to be hosted by the ISS. The second would be for a stand-alone satellite of \textbf{approximately 1 m\textsuperscript{3}}. The expected mass of the first concept is in the \textbf{5--10 kg range}, while the stand-alone mission would be around the \textbf{150 kg range} and would fall into an \textbf{ESA Mini-F / F mission class}. Based on the results of the MAQRO-PF mission, the full MAQRO mission would then aim for a larger \textbf{ESA M mission class}.
\subsection*{Floating Optical Breadboard}
\begin{figure}
    \centering
    \includegraphics[width=0.9\linewidth]{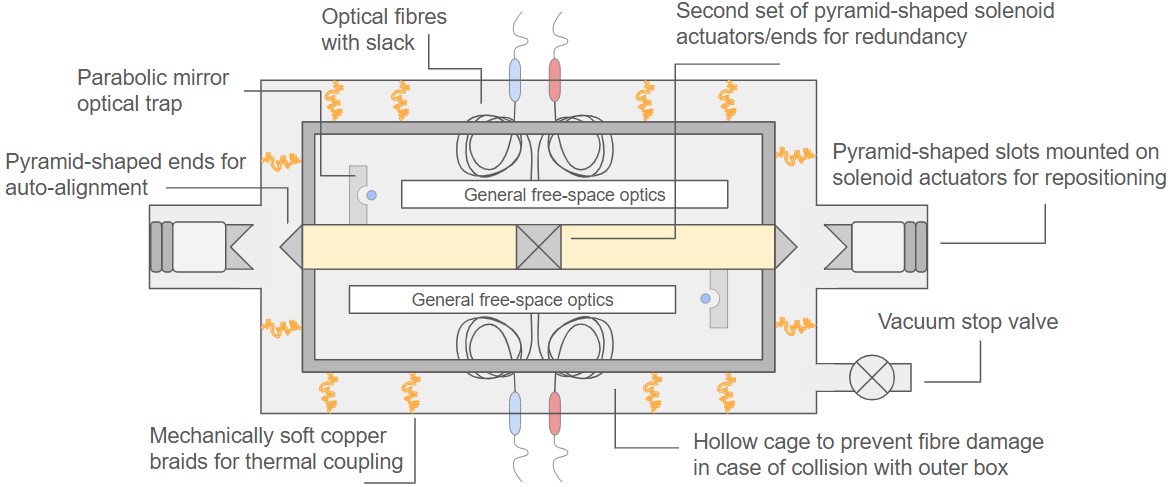}
    \caption{\centering Diagram of the design concept for the floating optical bench.}
    \label{FOB diagram}
\end{figure}

The heart of either mission concept is a 200 mm diameter vacuum chamber that would contain the experiment on a Zerodur `floating optical breadboard' (FOB). The FOB would be double-sided, with an identical copy of the optical trapping experiment on both sides such that the two experiments could be run simultaneously (see Figure \ref{FOB diagram}). We would aim to use a solenoid system, similar to the LISA-PF hold-and-release system that held the test masses in place during launch \cite{LISAPF-hold-and-release}, to release and recapture the FOB for each experimental run. The FOB is designed to decouple the payload from the vibration environment of the external satellite. Furthermore, the FOB would facilitate `drag-free' motion for the payload, partially removing the influence of orbital altitude drop resulting from atmospheric drag and solar radiation pressure which would otherwise significantly limit on experimental time. By designing pyramidal structures on the Zerodur and mating pyramidal sockets on the ends of the retractable solenoids, the FOB would be self-centring, returning to an equilibrium point from which it can be released again. A large part of the payload development would focus on perfecting the resilience of the solenoids, as the FOB would have to be released and recaptured up to hundreds of thousands of times over multiple years. One plan for improving their resilience is to include a redundant solenoid pair as a backup.
\subsection*{Laser Diffraction Grating}
A key component in any quantum matter-wave interferometer is a grating, or a series of slits, that must be placed in the path of the particles. This grating reveals the particle's quantum behaviour by causing the particle to `diffract'–-a situation that is only possible if the particle is existing as a quantum wave. The pattern of where the particles are recaptured after they are passed through the grating can be compared with classical and quantum theoretical predictions to assess whether the particle has existed in a quantum state. In most experiments, the diffraction grating is a solid mask through which a beam of quantum particles is passed. However, for our experiment, an optical grating is more suitable. We form an optical grating by reflecting a pulsed ultraviolet laser across the trapping region to create a standing wave grating, where the light forms the grid through which the particle must pass. The grating's position must remain stable to within a few nanometres for each experimental procedure such that we can obtain useable data. We must therefore mitigate any thermal expansion of the experiment. The primary mitigation technique is to use Zerodur as the experimental platform, as its thermal expansion coefficient is negligible. The FOB system also aids this as it will be thermally decoupled from the external satellite such that it should not be significantly impacted by the satellite's thermal environment.
\subsection*{Electronic and Thermal Stability}
While the FOB will be strongly thermally decoupled from the satellite, the payload's electronics, primarily the laser driver board, will require thermal stability to provide stable laser power to the experiment. During the development of the Op-to-Space payload, the laser driver board was tested across a 0\textdegree C--40\textdegree C range range and demonstrated full operation throughout. However, further thermal instability could cause the laser system to become unstable or inoperable. We have therefore set the payload's \textbf{thermal range to 0\textdegree C--40\textdegree C}. Additionally, to prevent any possible changes to the experiments during experimental procedures, we have set the limit on the rate of thermal change at \textbf{0.1\textdegree C / minute}.
\subsection*{Vacuum}
Another key aspect of our experiment is that it must be run in a vacuum so that gas molecules do not break down the quantum state. While conducting the experiment in the vacuum of space may seem like a simple solution, this is often not sufficient or stable enough for our experiments. Our experiment requires a \textbf{vacuum of at least 10\textsuperscript{-9} mbar}, with a \textbf{target of 10\textsuperscript{-13} mbar}. By comparison, public data from Robert A. Braeunig \cite{Braeunig} indicated that the vacuum in a 700 km orbit ranged from 2.1$\times10^{-8}$ mbar to 6$\times10^{-11}$ mbar, varying largely with solar activity. Further vacuum degradation will also occur due to out-gassing from the satellite and from gases released by the satellite's propulsion systems. We therefore plan to place the FOB inside a vacuum chamber that will be pumped to 10$^{-9}$ mbar before launch and also internally coated with `getter' material that will further remove gas molecules from the chamber, thereby improving the vacuum quality as time progresses. This technique was perfected for use in the Op-to-Space mission by Aquark, a spin-out of the University of Southampton.\\\\
To understand the impact of the vacuum conditions on our experiments, we must also be able to measure the vacuum around the experiment. Previous space missions have used either off-the-shelf or custom vacuum measurement systems, but none have demonstrated sufficient sensitivity to track pressure down to the $10^{-13}$ mbar target. A compact and low-power vacuum pressure gauge must therefore be designed and developed for this mission. One proposed option \cite{levitas} would also use the trapped particle to measure the surrounding pressure, but this technique needs significant development.
\subsection*{Optical Trapping Laser}
A core component of our experiment is the trapping laser system, which is required for capturing and manipulating the particle in the optical trap. In the Op-to-Space mission, the laser driver board was designed by and worked on in collaboration with Twin Paradox Labs, a UK spin-out of Surrey Space Centre. The board included an FPGA that enabled data acquisition and direct laser intensity modulation to control our particles. Twin Paradox Labs have since improved their board designs to provide greater functionality and reduced power consumption, making it ideal for further missions. However, a better understanding of the laser's lifetime is required, both from the standard laser diode ageing and through laser diode degradation due to the radiation environment of space (radiation darkening). Current plans include running two experiments simultaneously, each with their own trapping laser, with a second laser for each experiment kept in cold redundancy.
\subsection*{Data Communication}
Our experiments should run almost autonomously with minimal tracking and input from ground stations. To this end, the expected \textbf{standard command uplink} to the payload (following LEOP) will be in the \textbf{kilobit per day} range. Once the experiments are running, experimental data will only be collected for short periods before and after the particle and FOB are released/recaptured. We predict that, per experimental run, this would amount to 0.2 MB. We also expect that we would run approximately 300 experimental procedures per day, resulting in a \textbf{total expected data downlink of 60 MB / day}.

\section{Ongoing Development}
\subsection*{Loading Particles into the Optical Trap}
While our proposed mission will build on previous missions such as the Op-to-Space payload and LISA-PF, there are still many aspects of the payload that need further development. One of the current biggest challenges for conducting optical levitation experiments in space is how to load the particles into the trap. Most of the methods that are used in terrestrial experiments would not be suitable for a space platform due to excessive power consumption, the large volume of the hardware, or the vacuum degradation that would result from the process. The only currently developed loading method that would be suitable involves using piezoelectric crystals to `kick' the particles off a surface towards the trap. Variations of this method have been previously demonstrated for terrestrial experiments, but significant modifications are needed to reduce the power required for the technique, and to ensure that the particles travel at a sufficiently low speed for them to be captured in the trap, or conversely that they can be slowed sufficiently by the trap. Work is underway at the University of Swansea in partnership with the UKSA as part of the LOTIS project to advance this technique \cite{LOTIS}.
\subsection*{Autonomous Particle Control}
We must also develop the code architecture that will allow us to autonomously characterize, motionally cool to the ground state and state expand the particle, before selectively releasing it for the experimental procedure. Each of these procedures have been demonstrated separately in terrestrial experiments where human input is available. However, combining them into a single sequence that can run without input has not yet been achieved, and will take significant effort to produce. The computing element of the laser driver boards produced by Twin Paradox Labs should be sufficient to run all these processes once they have been perfected.
\subsection*{Laser Grating}
Another aspect of our payload that must be developed is the generation of a pulsed laser grating to interact with the particle while it is in its quantum state and cause it to diffract. Most off-the-shelf pulsed lasers of the required optical power and coherence that are used for terrestrial experiments would be unsuitable for space due to their large form factors and power consumptions. However, investigations were started during the development of the Op-to-Space payload, and a 405 nm laser and driver board were found that would produce laser pulses at the required rate. Further tests of the laser's coherence must still be undertaken, but integrating this laser into the experiment will then be trivial.

\section{Current Best Estimates}
\label{Sec_4}
The current best estimates (CBEs) of the platform requirements imposed by the proposed experiment are stated in Table \ref{CBEs} in order of importance to the experiment.
\begin{table}[h]
    \caption{Table of current best estimates of the payload's requirements.}    
    \centering
    \begin{tabular}{cc}
            \hline
            Requirement & CBE value \\  
            \hline\hline\\[-2.5ex]
            Linear Vibrational Stability & Minimum $10^{-7}$ g, Target $10^{-9}$ g\\ 
            Rotational Stability & 0.1 mRad / s \\
            Pointing Accuracy & 1 degree \\
            Pointing Knowledge & 0.1 mRad \\
            Thermal Stability & [0,40] \textdegree C, 0.1 \textdegree C / min \\
            Payload Power Consumption & 30 W \\
            Command Uplink & Kilobits, 1 set / day \\
            Telemetry Downlink & 60 MB / day \\
            \hline
            \end{tabular}

    \label{CBEs}
\end{table}

\section{Conclusion}
We have presented our plans for building a satellite-based optical levitation experiment that will explore the high-mass limits of quantum mechanical systems. We have discussed the requirements of our proposed payload and the various design elements that have been included to mitigate the limitations that the experiment will face. This white paper has built on the science objectives of the MAQRO mission proposals, and has demonstrated how it aligns with the UK space strategy.

\section*{Acknowledgments}
We thank Richard Robinson, Kathryn Graham, Simon Fellowes and Samantha Youles for their contributions to these plans during the CDF study conducted at the University of Portsmouth.
We also acknowledge funding by UKRI EPSRC (EP/W007444/1, EP/V000624/1 and EP/X009491/1), from the QuantERA II Programme (project LEMAQUME), grant agreement no. 101017733, the EU Horizon Europe EIC Pathfinder project QuCoM (101046973), and the Leverhulme Trust (RPG-2022-57).

\subsection*{Data Availability}
No new data was produced for this paper.

\begin{CJK}{UTF8}{gbsn}
\printbibliography
\end{CJK}
\end{document}